# Phase-decomposition-related short-range ordering in a Fe-Cr alloy

S. M. Dubiel[*] and J. Żukrowski

Faculty of Physics and Applied Computer Science, AGH University of Science and Technology, al. A. Mickiewicza 30, PL-30-059 Kraków

Abstract

A redistribution of Cr atoms related to a phase decomposition (PD) caused by an isothermal annealing at 415 $^oC$ in a 15.15 at%Cr Fe-Cr alloy was studied in an *ex-situ* way by the conversion electrons Mössbauer spectroscopy. Analysis of the spectra in terms of a two-shell model enabled determination of probabilities of 17 different atomic configurations and average numbers of Cr atoms within the first (*1NN*) and the second (*2NN*) neighbor shells versus annealing time, separately. The annealing-time evolution of these numbers, expressed in terms of the Cowley-Warren short-range order (SRO) parameters, was shown to follow the Johnson-Mehl-Avrami-Kolgomorov equation. The SRO-parameter averaged over the *1NN-2NN* shells was revealed to be linearly correlated with the average hyperfine field. Signatures in favor of the nucleation and growth mechanisms responsible for PD are discussed, too.

# 1. Introduction

Fe-Cr alloys have been exceptional both for scientific and industrial reasons. Concerning the former they show a wide spectrum of interesting properties including crystallographic and magnetic ones. Consequently, they have been regarded as good model alloys for testing various models and theories. The industrial importance of the alloys stems among other from their low ductile-to-brittle transition temperatures, strong resistance to swelling and bubble promotion as well as to high-temperature corrosion. Thanks to these properties, Fe-Cr alloys have been used as the basic ingredient in the steel making industry. Fe-Cr-based steels are regarded as good candidates for the design of various structural components in advanced nuclear energy installations such as Generation IV and fusion reactors [1]. In these circumstances, both experimental and theoretical studies of the alloys have been recently intensified [2-14].

At elevated temperatures, however, high-chromium steels are not stable. Two phenomena, having their origin in the crystallographic phase diagram of the Fe-Cr alloys, viz. (1) the phase decomposition (or phase separation) into the Fe-rich ($\alpha$) and the Cr-rich ($\alpha'$) phases at $T < \sim 500^oC$ leads to the so-called miscibility gap, and (2) the precipitation of the $\sigma$-phase that at $T > \sim 500^oC$ occur [10]. Both phenomena cause an enhancement of embrittlement, hence a severe deterioration of the mechanical materials properties. An interest in (1) is fourfold: (a) underlying mechanism(s), (b) borders of the miscibility gap i.e. the composition of $\alpha$ and $\alpha'$ phases, (c) kinetics of the decomposition, and (d) short-range ordering (SRO). Concerning (a), two different mechanisms viz. nucleation and growth (NG) and spinodal decomposition have been proposed [15]. According to calculations [10-12,16], the spinodal decomposition should be responsible for the decomposition of the alloys with Cr content, $x$, within the inner part of the miscibility gap viz. $\sim 0.2$-$0.3 \leq x \leq \sim 0.75$-$0.8$, while the nucleation and growth for those with $x < \sim 0.2$-$0.3$ and $x > \sim 0.75$-$0.8$. The existence of the miscibility gap has been experimentally confirmed by using different techniques among which small-angle neutron scattering (SANS), Mössbauer spectroscopy and atom probe analysis played the key role [17]. Yet, there is no unique evidence in favor of the exact position of the spinodal lines and the upper temperature limit of its existence. In particular, De Nys and Gielen revealed for $x = 20$ that aging at 540 $^oC$ resulted in the decomposition process via the nucleation and growth, while aging at 470$^oC$ via the spinodal decomposition [18]. The borders of the miscibility gap are still not known with a good enough precision. This statement is true for both theoretical and experimental results. The former differ quite significantly, depending on the model used in the calculations [10-12, and references therein], the latter are missing below $T \approx 400^oC$, and those above 400$^oC$ are scattered to such degree that any reasonable distinction between various predictions can be made [12]. The kinetics of the phase separation was studied by several workers, especially within the spinodal mode domain e. g. [19, 20], yet it cannot be regarded as definitely closed issue. Thorough knowledge of the activation energy, $E_a$, is still missing. In particular it is not known whether or not $E_a$ depends on the mechanism of the decomposition. According to available experimental data $E_a$=193-230 kJ/mole for the spinodal process [17] whereas 122 kJ/mole was found for the nucleation and growth [21]. It must be, however, remembered that the actual value of $E_a$ significantly depends on the chemical composition of alloys [22] as well as on the dislocations density [23]. Finally, concerning the phase decomposition-related SRO it seems to be, to our best knowledge, an issue that has not been studied to-date. Consequently, further studies are justified. The results we have obtained on the EFDA (European Fusion Development Agreement) model Fe-Cr14wt.% alloy using the Mössbauer spectroscopic technique are reported in this paper.

## 2. Experimental

A model EFDA Fe-Cr15.15 at.% alloy was chosen as a sample to be investigated in this study. Its chemical composition, as revealed by the microprobe analysis, is displayed in Table I.

Table I
Chemical composition of the investigated sample.

| Cr [wt. pct.] | C [wt. ppm] | S [wt. ppm] | O [wt. ppm] | N [wt. ppm] | P [wt. ppm] |
|---|---|---|---|---|---|
| 14.25 | 4 | 6 | 5.5 | 5 | <10 |

The studied sample was in form of a ~30 µm thick foil obtained from the original rod by cold rolling a ~1 mm thick disc cut off the rod. The phase decomposition was brought about by an isothermal vacuum (p ≤ $10^{-6}$ hPa) annealing at 415 $^{o}$C in a quartz tube. After the annealing was completed the tube was removed from the furnace and cooled down to room temperature during about 30 minutes. Next the sample was placed in a standard conversion electron detector. The measurements were performed by means of Mössbauer spectroscopy that has already proved to be a suitable tool to study the SRO-related issues as shown recently elsewhere [5]. Spectral parameters, and in particular, the hyperfine field, *B*, are not only sensitive to a presence of Cr atoms in the first two neighbor shells, *1NN-2NN*, but they even enable making the clear distinction between Cr atoms situated in *1NN* and those in *2NN* [5]. Thanks to this property, a proper analysis of Mössbauer spectra yields an information on probabilities of various atomic configurations, *P(m,n)*, *m* being a number of Cr atoms in *1NN*, and *n* that in *2NN*. Knowing atomic configurations, *(m,n)* and the *P(m,n)*-values enables, in turn, determination of an average number of Cr atoms in *1NN*, $<m>=\Sigma m \cdot P(m,n)$, that in *2NN*, $<n>=\Sigma n \cdot P(m,n)$, as well as in shells, $<m+n> = <m>+<n>$. In turn, the knowledge of $<m>$, $<n>$ and $<m+n>$ found in that way combined with the corresponding quantities expected for the random distribution, $<m>_r=8x$, $<n>_r=6x$, and $<m+n>_r=14x$, permits determining the Cowley-Warren SRO-parameters based on the following equations:

$$\alpha_1 = 1 - \frac{\langle m \rangle}{\langle m_r \rangle} \qquad (1a)$$

$$\alpha_2 = 1 - \frac{\langle n \rangle}{\langle n_r \rangle} \qquad (1b)$$

$$\alpha_{12} = 1 - \frac{\langle m+n \rangle}{\langle m+n \rangle_r} \qquad (1c)$$

In the present case, the Mössbauer spectra, examples of which are shown in Fig. 1, were measured at room temperature by recording conversion electrons (CEMS mode), hence they contain information on the presurface zone of the samples whose thickness ≤0.3 µm [24]. Although the probed volume constitutes merely ~1% of the sample, the information obtained from the spectra should be, in the light of recent Monte Carlo simulations [25], regarded as relevant to a bulk behavior, because a distribution of Cr atoms is affected by the surface only within the first few monolayers (ML) below the surface whereas the probed layer encompasses more then 2000 ML.

The measured spectra were analyzed in terms of two procedures:

- (I) a two-shell model, assuming the effect of the presence of Cr atoms in the *1NN-2NN* vicinity of the $^{57}$Fe probe nuclei on the hyperfine field, *B,* and the isomer shift, *IS*, was additive i.e. $X(m,n) = X(0,0) + m\Delta X_1 + n\Delta X_2$, where *X=B or IS*, $\Delta X_k$ is a change of *X* due to one Cr atom situated in *1NN* (*k*=1) and in *2NN* (*k*=2). The total number of possible atomic configurations *(m,n)* is equal to 63, but for *x* = 15.15 at% most of them have vanishingly small probabilities, so 17 most probable (according to the binomial distribution) were selected to be included into the fitting procedure (their overall probability was > 0.99). The spectra have been successfully fitted assuming: (a) the Lorentzian shape of the lines, (b) the same relative ratio of the Clebsch-Gordan coefficients for the corresponding lines (C1, C2 and C2), and the same line widths (G1, G2 and G3) in all subspectra associated with the considered configurations for a given spectrum. The best-fit spectral parameters obtained with this procedure are displayed in Table II. The obtained values of the $\Delta X_{1,2}$ parameters are in a good agreement with the corresponding ones reported elsewhere [5,26].

- (II) a hyperfine field distribution (HFD) method that yielded HFD-curves, *p(B)*, assuming a linear correlation between the hyperfine field and the isomer shift as found experimentally elsewhere [26]. This approach is model independent and allows determining of average values of *B*, $\langle B \rangle = \int p(B) dB$. More details on the fitting procedure (I) can be found, for example, in [5,26], while on (II) in [27,28].

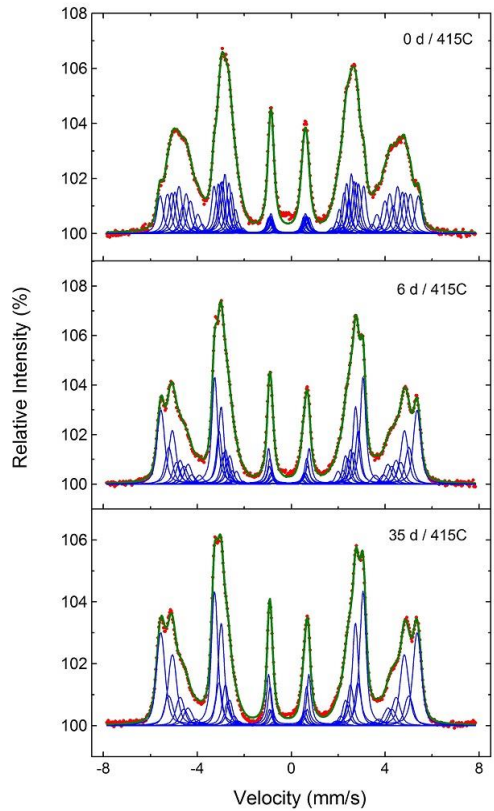

Fig. 1
(Online color) CEMS spectra recorded at RT on the Fe-Cr14 sample annealed at 415$^\circ$ C as labeled. Individual sub spectra corresponding to atomic configurations taken into account are indicated.

Table II
Best-fit spectral parameters as obtained with the fitting procedure (I). The ratio *C1/C3* was kept fixed at the value of 3. The meaning of the symbols is explained in the text. The value of *IS(0,0)* is given relative to the $^{57}$Co/Rh source. The values of G1, G2, G3 are half widths at half maximum.

| t [days] | B(0,0) [T] | $\Delta B_1$ [T] | $\Delta B_2$ [T] | IS(0,0) [mm/s] | $\Delta IS_1$ [mm/s] | $\Delta IS_2$ [mm/s] | G1 [mm/s] | G3 [mm/s] | G3 [mm/s] | C2/C3 |
|---|---|---|---|---|---|---|---|---|---|---|
| 0 | 33.95(4) | -3.24(6) | -2.10(4) | -.099(2) | - 021(1) | -.013(2) | .177(6) | .150(5) | .119(2) | 3.3(1) |
| 0.08 | 33.99(3) | -3.46(6) | -2.08(4) | -.101(2) | -.021(1) | -.012(2) | .167(4) | .149(4) | .122(2) | 3.6(1) |
| 0.17 | 33.82(2) | -3.07(5) | -1.99(3) | -.100(1) | -.020(1) | -.012(1) | .162(3) | .136(3) | .118(1) | 3.5(1) |
| 0.25 | 33.84(2) | -2.99(5) | -1.92(3) | -.103(1) | -.019(1) | -.013(2) | .166(4) | .135(4) | .118(2) | 3.2(1) |
| 0.33 | 33.83(2) | -3.12(4) | -2.00(3) | -.102(1) | -.020(1) | -.012(1) | .173(4) | .143(3) | .118(1) | 3.2(1) |
| 0.42 | 33.93(2) | -2.90(4) | -1.83(3) | -.104(1) | -.018(1) | -.011(2) | .160(4) | .134(3) | .115(2) | 3.6(1) |
| 0.5 | 33.89(2) | -3.04(6) | -1.95(4) | -.101(1) | -.0191) | -.014(1) | .171(4) | .133(3) | .118(2) | 3.4(1) |
| 0.54 | 33.92(2) | -3.11(5) | -1.98(3) | -.102(1) | -.0191) | -.012(2) | .165(4) | .136(3) | .119(2) | 3.5(1) |
| 1 | 34.05(2) | -3.18(4) | -2.01(3) | -.102(1) | -.020(1) | -.014(1) | .169(4) | .134(3) | .120(2) | 3.5(1) |
| 2 | 33.84(2) | -3.07(5) | -1.97(3) | -.105(1) | -.017(1) | -.011(1) | .165(3) | .133(3) | .116(2) | 3.6(1) |
| 3 | 33.74(1) | -3.06(3) | -1.97(3) | -.102(1) | -.020(1) | -.012(1) | .161(3) | .131(3) | .117(2) | 3.6(1) |
| 4 | 33.76(1) | -2.98(6) | -1.92(5) | -.102(1) | -.021(1) | -.012(1) | .169(3) | .134(2) | .118(1) | 3.5(1) |
| 6 | 33.79(1) | -3.15(4) | -2.02(3) | -.103(1) | -.022(2) | -.013(2) | .169(3) | .134(2) | .120(1) | 3.5(1) |
| 8 | 34.20(2) | -3.15(6) | -1.99(5) | -.105(1) | -.021(2) | -.013(2) | .191(4) | .147(3) | .122(2) | 3.4(1) |
| 10 | 33.87(2) | -3.07(5) | -1.92(4) | -.103(1) | -.019(2) | -.014(2) | .166(4) | .136(3) | .118(2) | 3.5(1) |
| 12 | 33.81(2) | -3.20(6) | -2.12(4) | -.103(1) | -.026(2) | -.010(2) | .171(3) | .136(2) | .119(2) | 3.5(1) |
| 16 | 33.90(2) | -3.09(5) | -1.91(5) | -.103(1) | -.021(2) | -.015(2) | .177(3) | .140(2) | .124(2) | 3.5(1) |
| 26 | 33.73(2) | -3.19(6) | -2.03(4) | -.104(1) | -.026(2) | -.010(2) | .175(4) | .138(3) | .121(2) | 3.3(1) |
| 35 | 33.91(2) | -3.23(6) | -2.06(5) | -.104(1) | -.022(2) | -.014(2) | .179(3) | .141(3) | .121(2) | 3.3(1) |

## 3. Results and Discussion

The *P(m,n)*-values of the six most probable configurations are presented in Figs. 2, 3 and 4. The corresponding values expected for the random distribution are marked by dashed lines. It is obvious that none of the six probabilities is random. *P(0,0), P(1,0), P(0,2)* and *P(1,0)* have values exceeding the corresponding random ones, while the values of *P(2,0)* and *P(1,1)* are smaller. Especially strong departure from randomness can be seen for *(0,0)* and *(0,2)*, in the former, as well as for *(2,0)* and *(1,1)*, in the latter. Such behavior can be interpreted in terms of clustering of Cr atoms, or an attractive potential between them. One can use the values of *P(0,0)* to estimate the concentration of Cr, *x*. In the random distribution approximation, $x = 1 - \sqrt{P(0,0)}$ yields *x*=13.8 at% at the start (real concentration is 15.15 at %) and 7.5 at% after 35 days of annealing. These figures give clear-cut evidence in favor of a Cr atoms clustering.

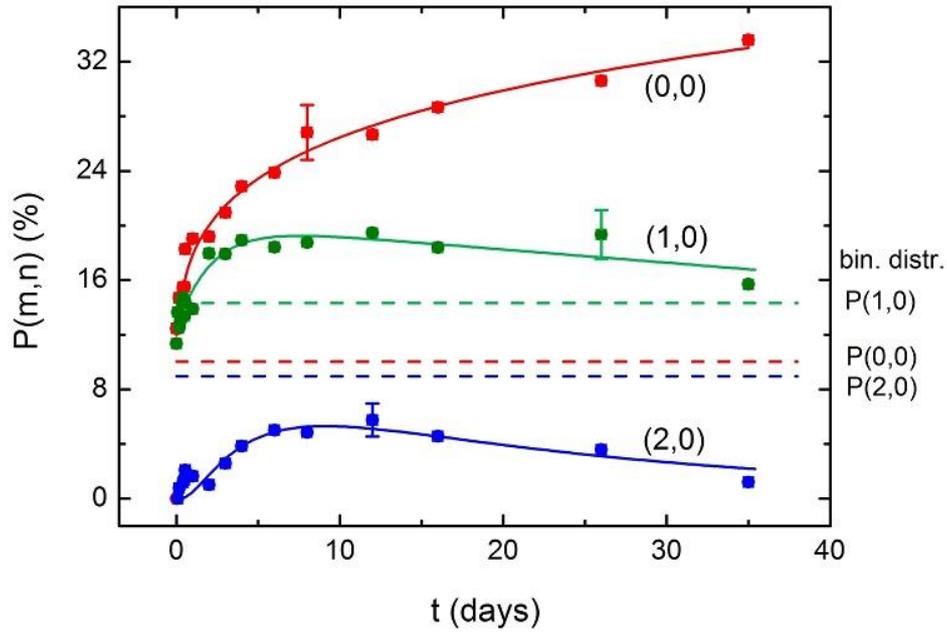

Fig. 2
(Online color) Probabilities of the atomic configurations (0,0), (1,0) and (2,0) as found by analyzing the spectra with the method (I) vs. annealing time, $t$. The solid lines are to guide the eye, while the dashed lines represent the behavior expected for the random distribution.

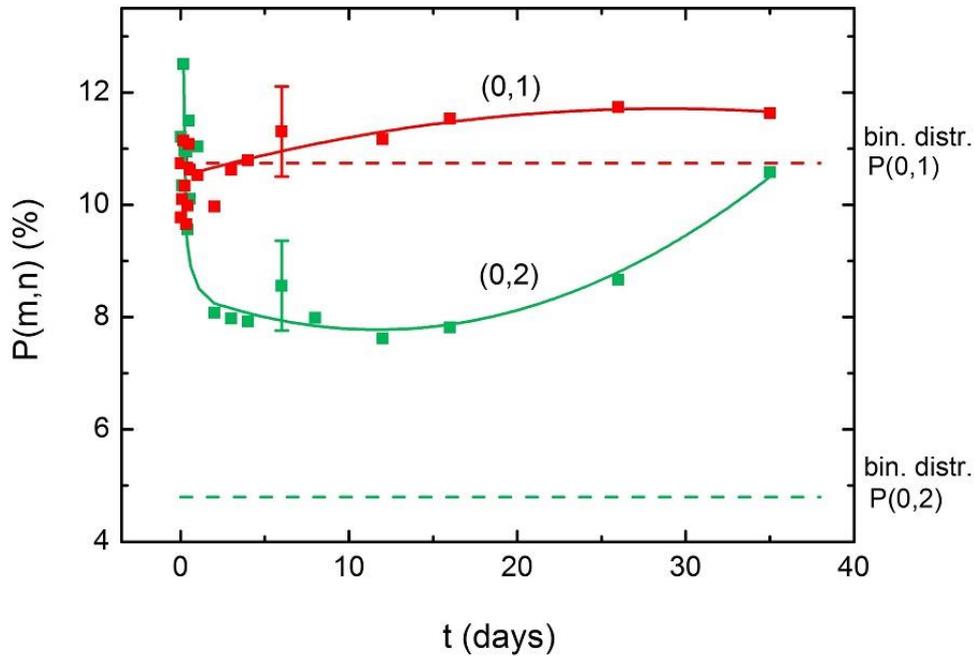

Fig. 3
(Online color) Probabilities of the atomic configurations (0,1) and (0,2) as found by analyzing the spectra with the method (I) vs. annealing time, $t$. The solid lines are to guide the eye, while the dashed lines represent the behavior expected for the random distribution.

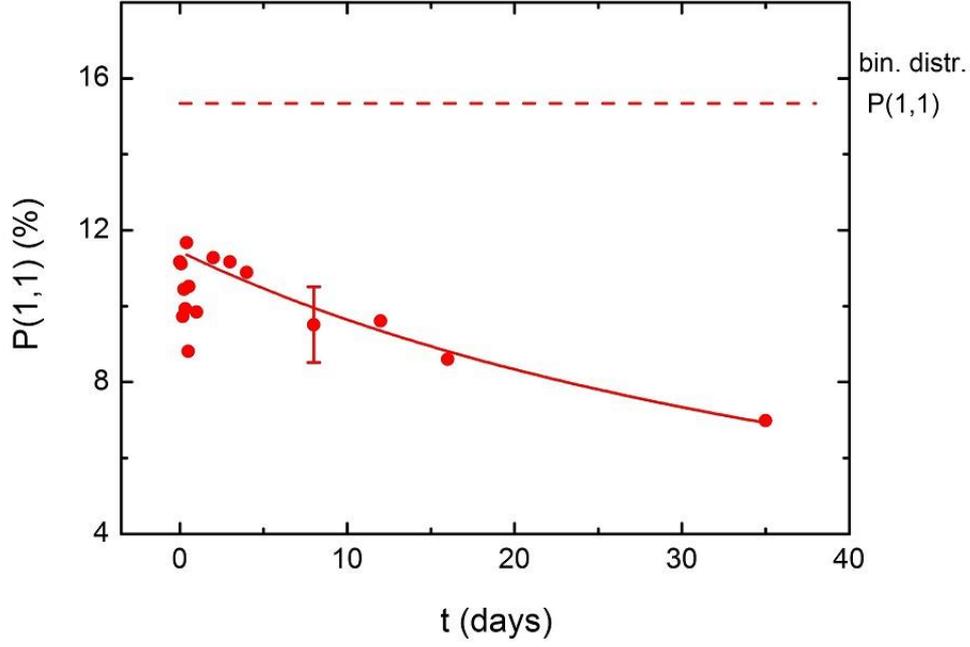

Fig. 4
(Online color) Probability of the atomic configuration (1,1) as found by analyzing the spectra with the method (I) vs. annealing time, *t*. The solid lines are to guide the eye, while the dashed line represents the behavior expected for the random distribution.

Knowing the probabilities of all 17 atomic configurations, that were taken into account, values of the SRO-parameters as obtained from the above-given equations are plotted in Fig. 5 versus the annealing time, *t*. Solid lines in Fig. 5 represent the best-fits to the data in terms of the Johnson-Mehl-Avrami-Kolmogorov (JMAK) equation:

$$\alpha_i = \alpha_{0i} + a\exp\left[-(kt)^n\right] \qquad (2)$$

Where *i*=1, 2, 12, *k* is the rate constant, and *n* is the Avrami exponent. Their values as found by fitting the JMAK equation to the data are displayed in Table III.

Table III
Best-fit kinetics parameters *k* and *n* as obtained by fitting the JMAK-equation to the SRO-parameters $\alpha_1$, $\alpha_2$ and $\alpha_{12}$, and to the average hyperfine field, *<B>*.

|  | k ($10^{-6}$ s$^{-1}$) | n |
|---|---|---|
| $\alpha_1$ | 6.4(7) | 0.6(1) |
| $\alpha_2$ | 5.1(9) | 0.3(2) |
| $\alpha_{12}$ | 6.0(7) | 0.6(1) |
| <B> | 6.0(9) | 0.6(1) |

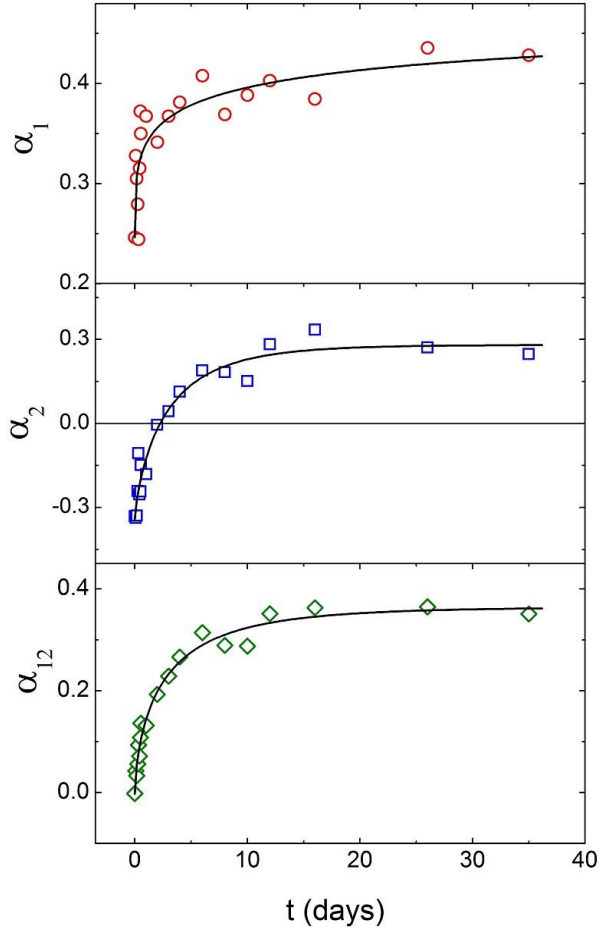

Fig. 5
(Online color) SRO-parameters as determined from equations (1a), (1b), and (1c) versus the annealing time, $t$. The solid lines represent the best fit to the data in terms of the JMAK-equation.

We note that in all three cases, the SRO-parameters could have been successfully fitted in terms of equation (2) what can be taken as evidence that the mechanism responsible for the phase decomposition in the investigated sample has a nucleation and growth character (contrary to the spinodal one which is alternative). Also the rate constant and the Avrami exponent are within the error limit the same for both shells. Concerning the values of the SRO-parameters, $\alpha_1$ starts with a positive value what means the actual number of Cr atoms in *1NN* shell is lower than the one expected for the random case. As the annealing proceeds, $\alpha_1$ further increases reaching the values of 0.43 in saturation. The latter feature can be taken as evidence that the phase decomposition has been practically completed i.e. a further annealing would not cause any statistically different distribution of Cr atoms. On the contrary, the initial value of $\alpha_2$ is negative i.e. the number of Cr atoms in the *2NN*-shell is higher than that expected for the random case. However, on annealing their number quickly decreases, and in the sample annealed for ~2.5 days $\alpha_2 = 0$. Longer annealing results in a positive value of $\alpha_2$ with a maximum of 0.3 in saturation. Concerning the average SRO-parameter, $\alpha_{12}$, its starting value was zero, and the final one 0.38. Positive values of all three SRO-parameters obtained after the phase decomposition has been practically completed mean that Cr atoms had formed clusters in the equilibrium state. The phenomenon is known as the *475°C embrittlement*, and it is responsible for the enhanced brittleness of Fe-Cr alloys and high-chromium steels.

The average hyperfine field, $<B>$, was shown to be linearly correlated with the average number of Cr atoms in the *1NN-2NN* shell [26]. Consequently, its time dependence during the phase decomposition should be describable with the JMAK-equation, and the kinetics parameters derived from the $<B>(t)$ dependence should be similar to those found for $\alpha_{12}$. Indeed, as illustrated in Fig. 6 and displayed in Table III, this is the case here. Consequently, there is a linear correlation between $<B>$ and $\alpha_{12}$ - see the inset in Fig.6.

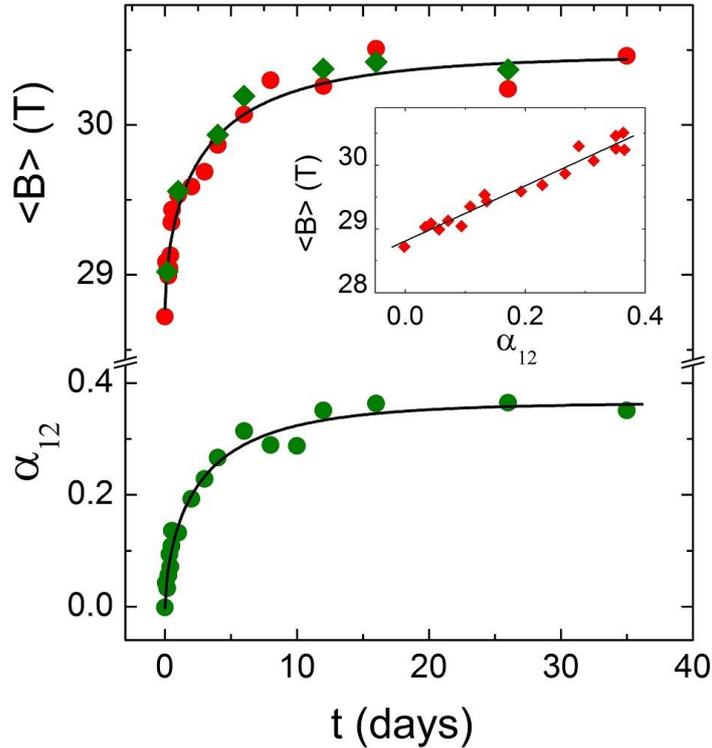

Fig. 6
(Online color) The average hyperfine field, $<B>$, and the average SRO-order parameter, $\alpha_{12}$ versus the annealing time, *t*. The solid lines in both instances stand for the best fits to the data in terms of the JMAK-equation. In the case of $<B>$ circles represent its values obtained with the superposition method while diamonds were found with the HFD method. The insert shows the $<B>$-$\alpha_{12}$ relationship with the best linear fit in form of a solid line.

The correlation is of a great practical value as it enables estimation of $\alpha_{12}$ from the knowledge of $<B>$. The latter can be easily determined by using model-free HFD methods in the analysis of the spectra [27,28].

The analysis of the spectra by means of the HFD-method gives us also a chance to get insight into the mechanisms responsible for the phase decomposition i.e. nucleation and growth or spinodal decomposition. The distinction between the two mechanisms can be easily made using Mössbauer spectroscopy. As already outlined by De Nys and Gielen [18], if the decomposition proceeds via the nucleation and growth, a paramagnetic subspectrum originating from the Cr-rich phase, α', should be present from the very beginning. However, as its amount depends on the concentration of chromium, its contribution for Fe-rich alloys may lie below the detection limit of the Mössbauer spectroscopy (~1% for a single-line spectrum). For example in the present case, its relative contribution should amount to ~2%. However, the detectability of a given phase by means of the used method depends not only on the abundance of the phase, but also on its recoilless factor, $f$, which is proportional to the concentration of Fe atoms. As the latter in α' is by a factor of ~7 smaller than in the Fe-rich phase, the contribution of α' must be correspondingly higher to be visible in the spectrum. Another signature that can be used for making the distinction between the two mechanisms is a behavior of a HFD histogram, hence the $p(B)$-curves. In the case of the nucleation and growth mechanism, the fragment of the HFD corresponding to the Fe-rich phase does not change its shape in the course of decomposition but its relative contribution grows at the cost of the HFD part representing the initial distribution [18,29]. A selection of the four $p(B)$-curves for three different annealing times and the one representing the initial distribution is displayed in Fig. 7. It is clear that the fragment of the HFD between ~30 and ~35 T does not change its two-peak shape as the phase decomposition goes on. Instead, the intensity of the two peaks increases at the cost of the peaks corresponding to $B < 30$ T. Such behavior is compatible with the nucleation and growth mechanism.

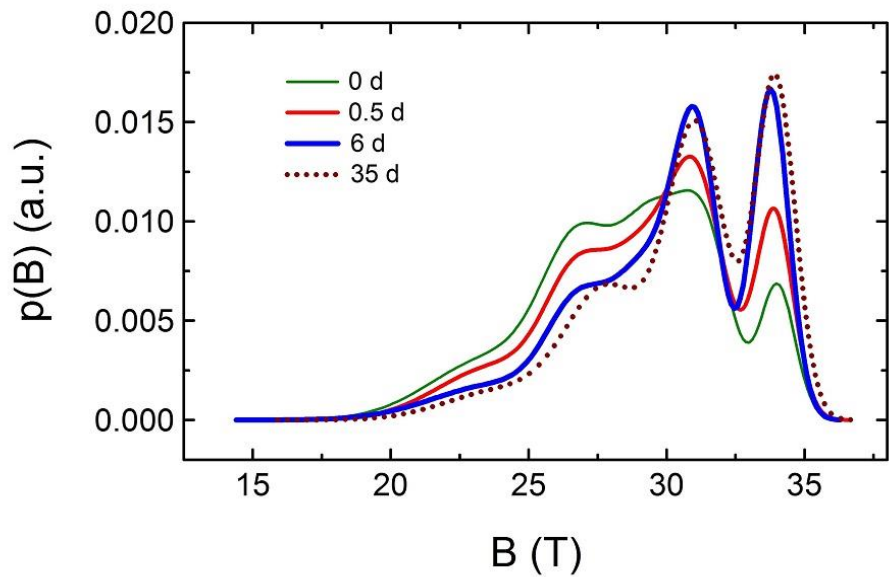

Fig. 7
(online color) $p(B)$-curves as obtained from the spectra analyzed with the HFD method for different annealing periods shown.

The peak in HFD located at ~33T can be ascribed to those Fe atoms that in their neighborhood do not have Cr atoms. Thus, an increase of its intensity with $t$ can be regarded as an indication of the decomposition process, hence can be compared with the $P(0,0)$-values obtained from the fitting procedure (I). To get the relative intensity of the peak, the $P(B)$-curves were fitted to five Gaussian-shaped distributions which are, for the sake of clarity, not shown.
A comparison between the $P(0,0)$-values and the relative intensity of the 33T peak is made in Fig. 8. A general good agreement can be seen, proving thereby that the interpretation outlined above is correct.

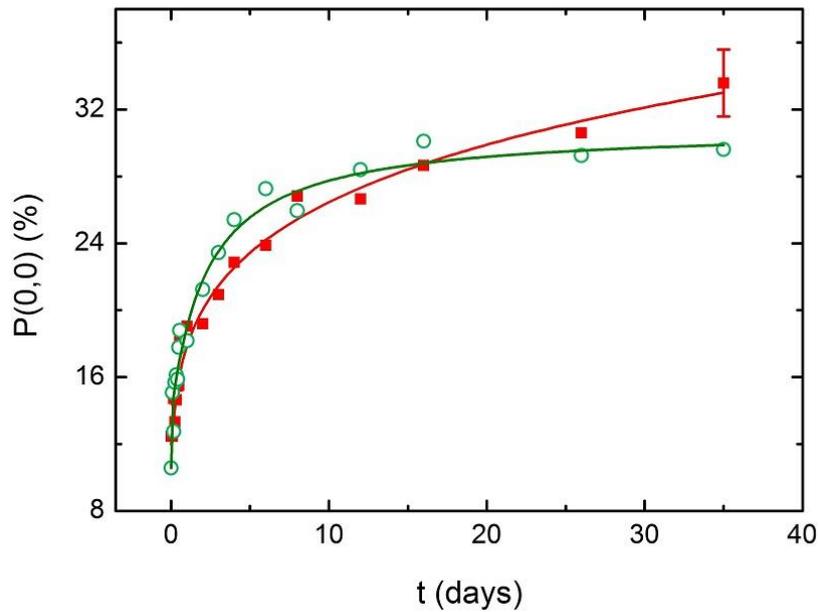

Fig. 8
(Online color) Probability of the (0,0) atomic configuration (squares) as found by analyzing the spectra with the method (I), and the relative contribution of the 33T peak (circles) as obtained using the method (II).

### 3. Summary

In summary, a redistribution of Cr atoms within the *1NN* and *2NN* neighbor shells brought about by an isothermal annealing at 415°C of the 15.15 at% Cr Fe-Cr alloy was studied with the CEMS method and expressed in terms of the Cowley-Warren short-range order (SRO) parameters. Evidence was found that the average number of Cr atoms decreases in both shells when the annealing proceeds, and eventually saturates what can be taken as evidence that the decomposition process has practically completed within the annealing period applied in this study. Such behavior means that Cr atoms formed clusters upon the annealing. The kinetics of the process could have been well described in terms of the JMAK-equation, proving thereby that the underlying mechanism had the nucleation and growth character. The kinetics parameters i.e. the rate constant, $k$, and the Avrami exponent, $n$, found for the average SRO-parameter, $\alpha_{12}$, were the same as those revealed for the average hyperfine field, $<B>$. Consequently, the two quantities were revealed to be linearly correlated. The latter finding is of a great practical importance as it enables a precise and easy determination of $\alpha_{12}$ from the knowledge of $<B>$. Changes in the *P(B)*-curves supporting the nucleation and growth mechanism were discussed, too.

### Acknowledgements

The study was supported by the EFDA-EURATOM-IPLLM association and the Ministry of Science and Higher Education (MNiSW), Warszawa.

* Corresponding author: Stanislaw.Dubiel@fis.agh.edu.pl